# Nanowire Lasers

C. Couteau[1,2,3], A. Larrue[1,4], C. Wilhelm[1], C. Soci[1,2]

[1]*CINTRA CNRS-NTU-Thales, UMI 3288, Nanyang Technological University, Singapore*

[2]*Centre for Disruptive Photonic Technologies (CDPT), Nanyang Technological University, Singapore*

[3]*Laboratory for Nanotechnology, Instrumentation and Optics, Charles Delaunay Institute, CNRS UMR6281, University of Technology of Troyes, France*

[4]*LAAS CNRS and Université de Toulouse, Toulouse, France*

We review principles and trends in the use of semiconductor nanowires (NWs) as gain media for stimulated emission and lasing. Semiconductor nanowires have recently been widely studied for use in integrated optoelectronic devices, such as LEDs, solar cells, and transistors. Intensive research has also been conducted on the use of nanowires for sub-wavelength laser systems that take advantage of their quasi-one-dimensional nature, flexibility in material choice and combination, and intrinsic optoelectronic properties.

First, we provide an overview on using quasi-one-dimensional nanowire systems to realize sub-wavelength lasers with efficient, directional, and low-threshold emission. We then describe the state-of-the-art for nanowire lasers in terms of materials, geometry, and wavelength tunability. Next, we present the basics of lasing in semiconductor nanowires, define the key parameters for stimulated emission, and introduce the properties of nanowires. We then review advanced nanowire laser designs from the literature. Finally, we present interesting perspectives for low-threshold nanoscale light sources and optical interconnects. We intend to illustrate the potential of nanolasers in many applications, such as nanophotonic devices that integrate electronics and photonics for next-generation optoelectronic devices. For instance, these building blocks for nanoscale photonics can be used for data storage and biomedical applications when coupled to on-chip characterization tools. These nanoscale monochromatic laser light sources promise breakthroughs in nanophotonics, as they can operate at room temperature, potentially be electrically driven, and yield a better understanding of intrinsic nanomaterial properties and surface state effects in low-dimensional semiconductor systems.

# **Table of Contents**

# 1. Introduction

Nowadays, lasers are ubiquitous in science and technology as well as in everyday life: they are vital for communications, sensing, and metrology, and are widely used for biological imaging, local surgery, metal welding, consumer electronics, and so on. The success and widespread application of semiconductor lasers are mainly because they are much smaller, consume less power, and are far cheaper than any other kind of laser. Further applications and consequent benefits for wider society are expected from a new generation of even smaller, cheaper, and more energy-efficient devices. Nanolasers have emerged as a new class of miniaturized semiconductor laser that are potentially cost-effective and easier to integrate. They consist of sub-micron-size "wires" typically formed of metal oxides, II-VI, or III-V semiconductor alloys. Owing to the large difference in refractive index between the semiconductor and the embedding medium (usually free space or a low–refractive index dielectric material), the wire acts as both the optical gain medium and the optical cavity, thereby allowing lasing despite the limited volume of gain material. These structures are broadly referred to as *semiconductor nanowire lasers*, and have been the subject of intense research since their first demonstration in 2001 [Hua01].

Although current top-down fabrication technology can achieve comparable dimensions, semiconductor nanowires are preferably obtained via a self-assembly, bottom-up growth process, which requires fewer post-processing steps than conventional diode lasers. In addition, bottom-up growth can yield nearly defect-free, single-crystal heterostructures on virtually any kind of substrate, thus tremendously increasing the spatial design parameters. This unique combination of geometric and material properties is therefore very attractive for developing highly integrated laser sources that, besides optical data transmission, can be used for sensing, imaging, and many other applications [Par13]. Low-threshold operation is expected owing to the small size of the active region of the semiconductor nanowire laser. Moreover, nanowire gain spectra are typically much broader than corresponding quantum wells, thus allowing wider tunability.

In this article, we first review the state-of-the-art of nanowire lasers in terms of materials, geometric configurations, and spectral emission properties. We then describe the basics of lasing in semiconductor nanowires, recalling the key parameters that underlie the process of stimulated emission. Finally, after reviewing advanced nanowire laser designs reported in the literature, we present interesting perspectives on low-threshold nanoscale light sources and optical interconnects, including the latest developments in electrical pumping and proposals for low-threshold operation based on coupling nanowires with external microcavities.

Due to space and focus constraints, this review does not aim to provide an exhaustive account of the enormous amount of work on nanoscale lasers, for which we refer the reader to excellent prior reviews [Nin10,Nin12,Van11,Zim10]. With the overall goal of identifying a roadmap toward laser-emitting structures with sub-wavelength dimensions, we consider only nanowires where at least one dimension is smaller than the effective wavelength of the active material. Furthermore, we focus on dielectric semiconductor nanowires, excluding random lasing effects in nanowire arrays (an extrinsic effect related to random cavities rather than inherent material properties) and nanowire lasers based on organic materials, like polymers [O'C07] or metals [Nez10]. In particular, we do not address work related to surface plasmon amplification by stimulated emission of radiation (spaser) [Ber03,Nog09], as more appropriate reviews on this topic already exist [Sto09,Leo12,Zay13].

# 2. Types of nanowire lasers

Yang's group first observed stimulated emission in nanowires in 2001 at Berkeley University, with multiple ZnO nanowires [Hua01]. They quickly followed with reports on lasing in single ZnO and GaN nanowires [Joh01,Joh02]. These indicated similarities between exciton or electron-hole plasma (EHP) radiative recombination in single nanowires with reflecting end facets, and standard gain media between two cavity mirrors in conventional lasers. Research on nanolasers has since expanded rapidly, with hundreds of papers published on the subject by over a dozen groups worldwide (*e.g.*, UC Berkeley, Harvard University, University of Karlsruhe, NASA Ames Research Center, Laboratory for Photonics and Nanostructures of CNRS, Nanyang Technological University, City University of Hong Kong, Ghent University, UC San Diego, and Utrecht University).

   a. <u>Materials</u>

Lasing has been successfully demonstrated in a variety of semiconductor nanowire materials, with emission wavelengths ranging from 370 to 2200 nm (Figure 1 and Table 1). These include II-VI and III-V semiconductors and metal oxides, such as ZnO [Hua01, Joh01], GaN [Joh02], CdS [Aga05], CdSe [Li13], GaSb [Chi06], ZnS [Din04], $Cr^{2+}$ doped ZnSe [Fen13], graded $CdS_xSe_{1-x}$ [Lua12], InGasAs/GaAs [Che11], GaAs/GaAsP [Hua09], and InGaN/GaN [Wu11]. The achievable wide spectral range allows various possible applications for these nanolasers. In the near–UV range, such applications can be sensing, water treatment, or biological; in the visible range, potential applications include lighting, sensing, and biomedicine coupled to lab-on-a-chip tools. Finally, in the near-infrared region, applications are expected in telecommunications, where the scaling down of laser sources for future communications is critical.

The first observations were made in wide bandgap materials like ZnO [Hua01,Wu02,Yan02] and GaN [Joh02]. Quantum efficiency (QE) measurements of ZnO nanolasers yielded external QEs as high as 60% and internal QEs of 85% [Zha05]. This conversion is very efficient considering the high density of surface states in nanowires, which typically act as traps or non-radiative recombination centers. Stimulated emission due to two- and three-photon absorption with a high threshold of 100 $mJ/cm^2$ was also reported in ZnO nanowires [Zha06]. Zhang *et al.* further studied this and demonstrated a lower threshold of 160 $\mu J/cm^2$ from two-photon absorption of ZnO nanowires pumped with a pulsed laser at 700 nm [Zha09a]. Following these pioneering works, II-VI compounds became the focus of investigation, and Lieber's group at Harvard University first studied the lasing properties of CdS nanowires and the effect of temperature on the lasing threshold [Dua03]. Researchers at the City University of Hong Kong first measured lasing in ZnS nanowires (grown by electrochemistry) and single nanoribbons in 2004 [Zap04, Din04]. The latter had a threshold of 60 $kW/cm^2$. Even longer emission wavelengths were obtained in $Cr^{2+}$-doped II-VI nanowires, like a $Cr^{2+}$-doped ZnSe nanowire laser emitting at 2194 nm [Fen13].

Stimulated emission in III-V compound nanowires was first reported in GaSb with lasing emission around telecommunication wavelengths (1570 nm) [Chi06]. Chin *et al.* from the NASA Ames Research Center were the first to demonstrate low-temperature lasing emission of a III-V nanowire in the near-infrared, with $I_{th}$ = 50 $\mu J/cm^2$ [Chi06]. One of the greatest potentials of III-V nanowires is for their integration onto silicon. Chen *et al.* at the University of Berkeley demonstrated lasing from InGaAs/GaAs core/shell nanowires grown directly onto a silicon substrate [Che11]. Exhibiting emission around 900 nm for a lasing threshold of ~60 $\mu J/cm^2$ and a quality factor of 206, this result paved the way for integration of III-V emitters with Si photonic chips. In the near-infrared range, three recent reports demonstrated room-temperature lasing from InP integrated on silicon [Wan13b] and of GaAs nanowires with a AlGaAs shell [Sax13,May13].

b. Tunability

For future nanolaser applications, two key features must be considered: broadband emission and wavelength tunability. Different methods have been used to achieve tunability. The first is based on spatially graded alloy composition achieved during nanowire growth. This was demonstrated on a single chip (1.2-cm long) using the ternary alloy CdSSe, with emission wavelengths ranging from 500 to 700 nm [Pan09] at 77 K. Lua *et al.* extended this emission wavelength down to 340 nm with the ternary alloy ZnCdS [Lua12]. The second tunability method is based on the self-absorption/re-absorption mechanism (see Section 3-b). Li *et al.* achieved wide tuning of CdSe nanowire emission by shifting the pump spot position along the nanowire or by modifying nanowire length [Li13]. This tuned the lasing wavelength from 706 to 746 nm upon self-absorption of higher energy photons by the nanowire and subsequent reemission of lower energy photons, causing the emission wavelength to red shift. This method was also realized in CdS nanowires, where the lasing modes themselves could be similarly tailored [Liu13].

An alternative method for wavelength tunability consists of making a heterostructure in the same way one achieves tunable sources with quantum wells [Qia08] (see Section 4-b). Table 1 summarizes some key work in semiconductor nanolasers. Finally, Ding *et al.* proposed an interesting approach to broaden the range of emission wavelengths: they coated an optical micro-fiber with three different types of nanowires (ZnO, CdS, and CdSe), thus achieving lasing at 391, 519, and 743 nm simultaneously. Although the relatively low threshold of 1.3 µJ appeared promising, other limiting effects (*e.g.*, re-absorption) will likely limit application of this type of structure to white-light emission with no controlled tunability [Din09].

c. Geometry and dimensions

Nanolasers based on a variety of crystalline shapes and characteristic dimensions have been reported in the literature, such as nanowires [Hua01], nanoribbons [Yan03b], dendritic nanowires [Yan03a], and nanowire ring resonators [Pau06], with typical diameters ranging from 20 to 500 nm and lengths from 1 to 100 µm. Besides ZnO nanowires, Yang *et al.* demonstrated stimulated emission of nanoribbons [Yan03a] and dendritic arrays of nanowires [Yan03b]. ZnS nanoribbon lasers have also been reported [Zap04], although they appear less promising than nanowires owing to the higher threshold and the appearance of complex optical multimodes within the nanoribbon/nanobelt owing to the rectangular cross-section.

Although most theoretical models assume cylindrical nanowire cross-sections [Mas03,Mas04c,Che06,Zha08], which in most cases is a good approximation, semiconductors with cubic or hexagonal structures usually exhibit triangular (*e.g.*, GaN [Gra05,Seo08]) and hexagonal (*e.g.*, ZnO [Joh03,Nob05], or III-V nanowires [Hua09,Che11]) cross-sections.

Finally, there are still many challenges in controlling the growth of nanowires, in terms of length and diameter. In particular, it is still difficult to grow a small-diameter (< 20 nm) nanowire due to the onset of the Gibbs–Thomson effect.

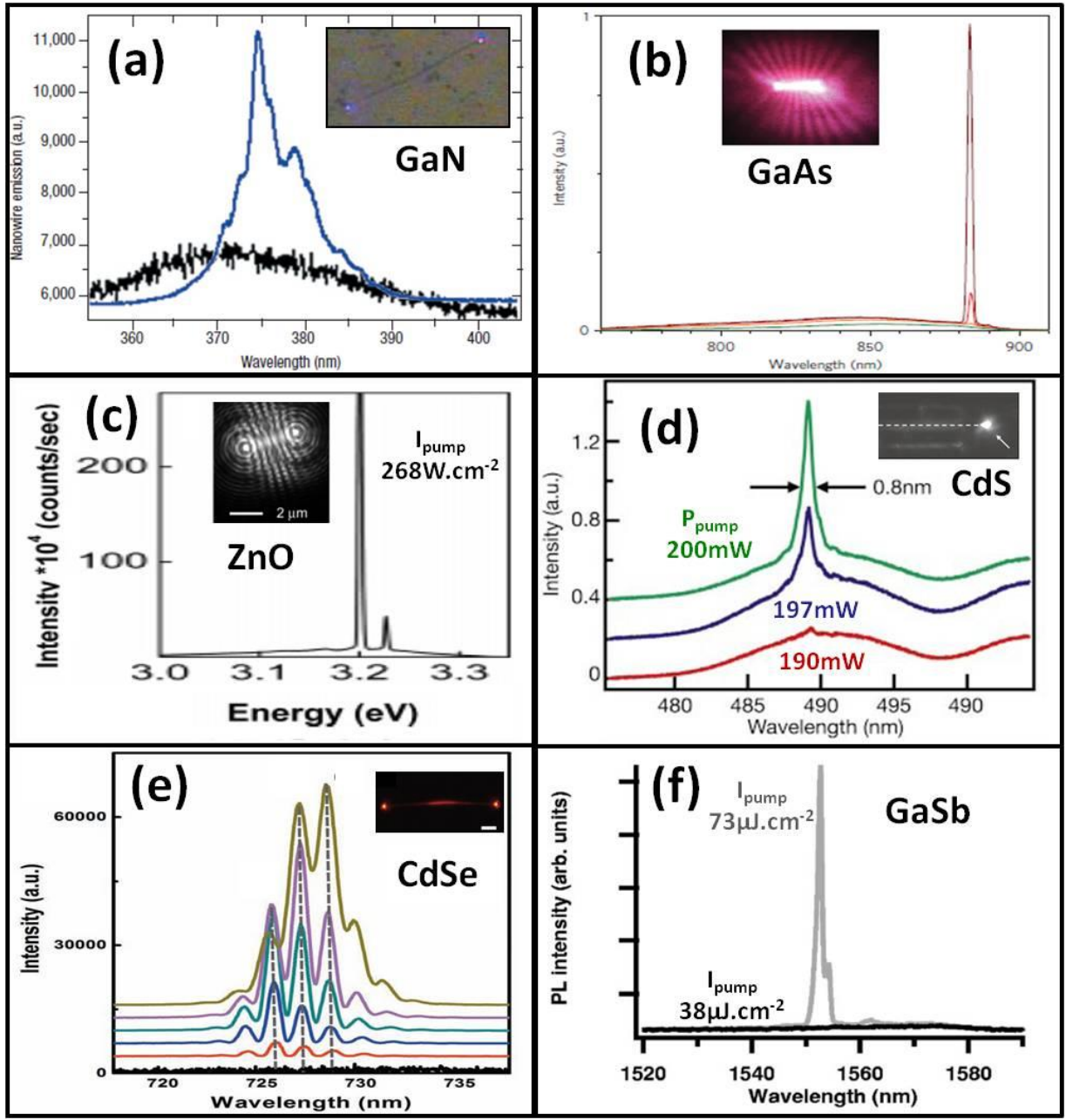

**Figure 1.** Semiconductor nanowire lasers of different materials. (a) GaN with a photoluminescence spectrum at 1-mW continuous wave excitation (black) and lasing with pulsed 1 nJ/cm$^2$ excitation (blue) (adapted with permission [Joh02]). (b) GaAs nanowire with transition spectra from spontaneous emission with 144 µJ/cm$^2$ per pulse (green) and 202 µJ/cm$^2$ per pulse (orange) to stimulated emission with 240 µJ/cm$^2$ per pulse (red) and 288 µJ/cm$^2$ per pulse (brown) (adapted with permission [Sax13]). (c) Stimulated emission spectra of a single ZnO nanowire (adapted with permission [vaV06]). (d) Transition spectra from spontaneous (red) to stimulated (blue and green) emission of an electrically driven CdS nanowire laser for different pump powers (adapted with permission [Dua03]). (e) Emission spectra of a CdSe nanowire by increasing pumping intensity from 69 to 547 µJ/cm$^2$ above the threshold (adapted with permission [Li13]). (f) GaSb transition spectra from spontaneous emission (black) to stimulated emission (gray) (adapted with permission [Chi06]). Insets (a–e) show optical microscope images of the lasing nanowires.

**Table 1. Summary of key works in semiconductor nanolasers since the first demonstration of lasing in nanowires in 2001** (VPT: Vapour Phase Transport, VLS: Vapour-Liquid Solid, LACG: Laser Assisted Catalytic Growth, EC: ElectroChemistry, MOCVD: Metal Organic Chemical Vapour Deposition, RVT: Reactive Vapor Transport, SA-MOPVE: Selected Area Metal Organic Vapour Phase Epitaxy, PE: Physical Evaporation, BC: Biocatalysts, MBE: Molecular Beam Epitaxy, PA-MBE: Plasma-Assisted MBE, PLGE: Photolithography Etching, LA: laser Ablation FPAL: Femtosecond Pulsed Laser Ablation in Liquid). (1) 4.2-300 K (2) 20-150 K (3) 10-50 K (4) 4.2-125 K (5) 77 K (6) 7 K.

| Ref. | Notes | Material | $\lambda$ (nm) | Q factor | Threshold | Diameter (nm) | Length (µm) | Growth | Single NW |
|---|---|---|---|---|---|---|---|---|---|
| [Hua01] | First report | ZnO | 385 | >50 | 40 kW/cm$^2$ | 70–100 | 2–10 | VPT | no |
| [Joh01] | First single NW | ZnO | 380 | ~130 | 120 kW/cm$^2$ | 140 | 5 | VPT | yes |
| [Joh02] | Lasing cavity effect | GaN | 375 | 500–1500 | 700 nJ/cm$^2$ | 300 | 40 | VLS | yes |
| [Dua03] | Electrical pumping | CdS | 512 | 600 | 40 kW/cm$^2$ | 140 | 18.8 | LACG | yes |
| [Yan03b] | First nanoribbons | ZnO | 386 | ~3000 | 3.2 µJ/cm$^2$ | 10–300 | 15 | VLS | yes |
| [Din04] | UV lasing | ZnS | 338 | ~1100 | 130 kW/cm$^2$ | 5–200 | 24 | EC | no |
| [Aga05] | Low T (1) | CdS | 490 | ~1600 | 0.2 µJ/cm$^2$ | 100 | 40 | VLS | yes |
| [Chi06] | Telecom wavelength (2) | GaSb | 1550 | NA | 50 µJ/cm$^2$ | 700–1500 | 35 | RVT | no |
| [Par07] | NW + microstadium resonator | GaN | 372 | 3500 | 1540 kW/cm$^2$ | 170 | 7.5 | MOCVD | yes |
| [Zho07] | Low T, ordered arrays of NWs (3) | ZnO | 371 | >3700 | 2600 µJ/cm$^2$ | 200 | 4.7 | VPT | no |
| [Qia08] | First heterostructure | GaN/(InGaN/GaN)n | 365–494 | 650–2500 | 600–2500 kW/cm$^2$ | ~400 | 50 | MOCVD | yes |
| [Hua09] | Near-visible (4) | GaAs/GaAsP | 815 | NA | 8.4 kW/cm$^2$ | 200–500 | 2–6 | SA-MOPVE | no |
| [Pan09] | Widely tunable 500–700 nm (5) | CdS$_{1-x}$Se$_x$ | 500–700 | NA | ~260 kW/cm$^2$ | 200 | 10–30 | PE | no |
| [Zha09a] | Two-photon pumping | ZnO | 388 | 1900 | 160 µJ/cm$^2$ | 180 | 10 | BC | no |
| [Che11b] | Single-mode, helical modes, on Si | InGaAs/GaAs | 950 | 206 | 93 µJ/cm$^2$ | ~500 | ~3 | MOCVD | yes |
| [Chu11] | Electrical injection in homojunction | ZnO | 390 | NA | 190 kW/cm$^2$ | 200 | 3.2 | MBE and CVD | no |
| [Wu11] | NWs onto plasmonic film (6) | InGaN/GaN | 530 | 100 | 300 kW/cm$^2$ | 30 | 0.68 | PA-MBE | no |
| [Li12] | Top-down approach | GaN | 370 | ~3000 | 231 kW/cm$^2$ | 135 | 4.7 | PLGE | yes |
| [Lua12] | Wide-range wavelength; same substrate | CdS$_{1-x}$Se$_x$/Zn$_y$Cd$_{1-y}$S | 340–710 | up to 3000 | 35–90 kW/cm$^2$ | NA | NA | LA | no |
| [Xu12a] | Single-mode lasing by Vernier effect | GaN | 370 | ~2600 | 875 kW/cm$^2$ | 700 | 7.8 | PLGE | yes |
| [Fen13] | Doped NWs and far-IR emission | Cr$^{2+}$ in ZnSe | 2000–2500 | NA | 340 µJ/cm$^2$ | 30–120 | 10–40 | FPLAL | no |

| [Li13] | Tunable by self-absorption method | CdSe | 706–746 | NA | 550 μJ/cm$^2$ | 400 | 0.29–8 | VLS | yes |

# 3. Lasing mechanisms in nanowires

In conventional semiconductor lasers, various gain media (such as multiple quantum wells or quantum dots) are typically constructed during the epitaxial growth of the entire vertical heterostructure, while the laser cavity is formed by conventional top-down micro/nano-fabrication techniques with multi-step processes. Thanks to their quasi-1D geometry, individual nanowire lasers—unlike their traditional counterparts—are able to merge the optical resonator with the gain medium. This remarkable feature changes the way semiconductor lasers are considered, transforming conventional multi-step top-down fabrication into one-step bottom-up growth, and most importantly allowing miniaturization and integration of light sources on heterogeneous materials. This opens unprecedented possibilities for optical resonator designs and establishes single nanowires as credible components for future integrated light sources and optical interconnects. In this section, we first discuss the fundamental electromagnetic properties of the propagative modes allowed in a single nanowire and for different materials. Next, we consider the bare resonator consisting of a single nanowire on a substrate. We then derive the gain coefficient for stimulated emission in semiconductors, especially in 1D systems such as nanowires. A typical lasing experiment is also described, and the relevant parameters for lasing emission in nanowires are highlighted. Finally, we discuss the inherent problems and sources of losses in such systems.

a. <u>Waveguiding mechanism and bare resonator</u>

The conception of nanowires as "ultimate" nanolasers began with the first experimental report on lasing emission characteristics under optical pumping of a single GaN freestanding nanowire removed from its original growth substrate and placed on a host substrate [Joh02]. This demonstrated that the nanowire can form a stand-alone Fabry–Pérot optical resonator and consequently highlighted one key prerequisite for laser cavity design: the sub-wavelength optical confinement of the nanowire geometry. Table 2 summarizes the basic parameters of laser physics. The main component is the Fabry–Pérot cavity consisting of a gain medium between two mirrors. We can then define parameters such as the quality factor, cavity gain, free spectral range, transmission of the Fabry–Pérot (FP) cavity, and the finesse of the cavity etc. (see Table 2). In the case of nanowires, the gain medium fills the entire cavity and waveguiding thus plays a major role.

Theoretical studies have been made on waveguiding and light-confinement mechanisms for a freestanding nanowire in air. Maslov and Ning extensively studied the electromagnetic characteristics of guided modes in single freestanding GaN, ZnO, and CdSe nanowires emitting in the visible range [Mas04a]. By resolving Maxwell's equations using a similar formalism to an infinite circular waveguide, they underlined the ability of the nanowire geometry to sustain multiple guided modes, depending on its diameter, with conventional electromagnetic symmetries, *i.e.*, hybrid modes with HE symmetries, and conventional transverse electric (TE) and transverse magnetic (TM) modes [Mas04b]. In particular, calculating the dispersion curves of the first guided modes ($HE_{11}$, $TE_{01}$, $TM_{01}$) yields an estimate of the efficiency of light confinement in such sub-wavelength waveguides. Indeed, these curves reveal two interesting regimes. The first occurs below the cut-off diameter of the first transverse mode ($TE_{01}$ mode), where only the fundamental $HE_{11}$ mode is confined inside the wire, enabling the waveguide to have a single mode along the transverse axis. However, this mode's effective index is not maximized, meaning that the electric field extends considerably outside the wire and hence generates significant propagation loss. The second regime lies above the cut-off frequency of the first transverse mode; at higher diameters, several guided modes are strongly confined inside the wire, with an effective index asymptotically approaching the semiconductor index, thus reducing the propagation losses of fundamental modes throughout the cavity. However, the presence of several guided modes increases the risk of the device operating in a multimode regime. Moreover, the various electric field distributions of confined modes that overlap with the gain medium will affect the device performance and favor mode competition via gain discrimination.

It is noted that the waveguiding mechanism sets the nanowire diameter towards that of a single nanowire laser. Figure 2-a presents the effective indices of the $HE_{11}$ (dashed) and $TE_{01}$ (solid) propagating modes versus diameter in infinite cylindrical nanowires for different semiconducting materials, with respect to their emission wavelength. Clear cut-off diameters are observed as we decrease the diameter. Despite the large refractive index mismatch between the nanowire's core and the surrounding medium (typically air), which favors strong confinement of the electric field, the diameter range for strong confinement of single or multiple guided modes has a minimum value, and is scaled by the refractive index and emission wavelength of the semiconductor. For instance, efficient waveguiding can be achieved for a 100-nm-diameter GaN nanowire emission and a minimum ~165-nm-diameter GaAs nanowire emission at 870 nm [Lar12].

Sub-wavelength optical confinement, attractive as it may be, is insufficient for obtaining a good-quality cavity. Consider a simple FP cavity formed by a freestanding nanowire in air, as depicted in Fig. 2-a (inset). The reflections at the end-facets of the nanowire are necessary for designing a laser source, as they are directly correlated to the threshold condition. However, computing these reflections is not as straightforward as in a conventional ridge laser, as the plane-wave approximation is no longer valid. The modal reflectivities at the end-facets of nanowires with circular or hexagonal cross-sections have been calculated with Finite- Difference Time Domain (FDTD); the reflectivity of the first TE mode was found to attain 80%, almost twice that of the fundamental $HE_{11}$ mode [Lar12]. This leads to mode competition and complicates the control of lasing.

Svendsen *et al.* reached similar conclusions after applying a semi-analytical approach based on reflection matrices to both wide bandgap materials and GaAs/AlGaAs core/shell nanowires [Sve11]. Wang *et al.* made a complete investigation into the reflectance and quality factor of the resonant modes as a function of the radius and length of a single nanowire with a dielectric constant close to that of ZnO, GaN, and CdS at the lasing frequencies, in the more realistic case of a nanowire standing on a substrate [Wan06]. Quality factors up to several hundred can be achieved for both the $HE_{11}$ and $TE_{01}$ modes in appropriately sized cavities. Analogous values for the absolute reflection coefficients of the different modes have been demonstrated. Three-dimensional FDTD calculations also show the impact of length on the Q factor of the resonant modes (Table 2), and cavity lengths >5 µm favor the creation of strong resonant modes [Wan06]. Recently, it has been demonstrated that a single nanowire cavity with a triangular cross-section lying on a $SiO_2$ substrate [Seo08] can also sustain high-Q FP resonance, with Q close to 500 for the TE-like mode.

Another key figure of merit for nanowire lasers is the confinement factor Γ of the resonant modes, as a large Γ will favor a low threshold (Table 2). Γ characterizes the overlap of the lasing mode with the gain medium, which for a nanowire depends strongly on its size and shape as well as on the electric field distribution of the resonant mode. For instance, a large confinement factor (Γ > 0.8) can be achieved in a nanowire with a sufficiently large triangular cross-section for the $TE_{01}$ mode [Seo08]. We note that producing a large confinement factor is only meaningful when the polarization of the electromagnetic guided mode matches the polarization of the dipole transition of the semiconductor nanostructure. Both the crystalline structure (wurtzite and zincblende) and the dielectric mismatch between the wires have been demonstrated to cause optical anisotropy, which greatly impacts the polarization of the light emitted from the wire [Wil12].

b. Amplified spontaneous emission and lasing

At this stage, it is useful to recall a typical lasing experiment in semiconductors. Since the mid-60s, lasing has been achieved in thin films and quantum wells in bulk or 2D semiconductor structures. If no particular modifications are made (such as sandwiching the thin film between two Bragg mirrors), then gain can be observed by pumping the semiconductor thin film with a pulsed laser. Stimulated emission is achieved at a certain threshold and two features are typically seen. First, the emission of the material as a function of pump laser power will encounter a drastic change of slope at a certain threshold. Second, one usually observes line narrowing in the emission spectrum, again indicating a transition from spontaneous to stimulated emission. Other indicators exist, including the reduction of carrier lifetime. The first step with semiconductor lasers is generally to achieve optical pumping before electrical pumping, which is usually harder as it requires precise doping of the material. Nanowire lasers have the same indicators of lasing. Nevertheless, it is interesting to see which physical mechanisms are involved in semiconductors as opposed to dye lasers or gas lasers, where the medium is normally simpler as all the emitters are identical atoms or molecules.

In semiconductors, stimulated emission occurs when the Bernard–Durrafourg condition is achieved. This occurs naturally when one examines the dipolar interaction between an electromagnetic wave and a semiconductor with a direct bandgap and assumes vertical optical transitions occur. The standard dipolar Hamiltonian leads to coupling between the different energy states of different bands within the semiconductor, resulting in optical transitions and photon emission. We define the optical susceptibility $\chi_{\vec{k}}(\omega)$ associated with the transition between the quasi-discrete levels $E_v(\vec{k})$ from the valence band and $E_c(\vec{k})$ from the conduction band. $\chi_{\vec{k}}(\omega)$ depends on $\rho_j(\omega)$, the joint density of states as well as $f_c(E)$ and $f_v(E)$, Fermi occupancy probabilities for the conduction and valence bands respectively in quasi-thermal equilibrium. This quasi-thermal equilibrium is necessary for achieving stimulated emission via strong non-equilibrium carrier densities of electrons and holes. We recall that the density of states depends on the geometry of the structure, which is constant for 2D materials and inversely proportional to the square-root of the energy in the case of 1D nanowires. This leads to van Hove singularities, where resonant photoabsorption can occur without the need for excitons. If anything, this should support lasing in 1D nanowires as opposed to the density of states of 2D materials, which do not possess such singularities [Arn13]. However, even a 1D system like a nanowire would be considered bulk if its diameter is relatively larger than the Bohr radius of the carriers (more than several tens of nanometers).

The absorption coefficient can now be derived, as it is related to the imaginary part of the total susceptibility and is given by:

$$\alpha(\omega) = \frac{\omega}{c.n_{opt}} \chi_{Im} = \frac{\pi q^2 x_{vc}^2 \omega}{\varepsilon_0 \hbar.c.n_{opt}} \rho_j(\omega)[f_v(\hbar\omega) - f_c(\hbar\omega)] = -\gamma(\omega) = \alpha_0(\omega)[f_v(\hbar\omega) - f_c(\hbar\omega)],$$

with $n_{opt}$ being the optical refractive index of the medium and $\gamma(\omega)$ the gain coefficient. Thus, the condition for optical gain is given by $\alpha(\omega) < 0$, which leads to the Bernard–Durrafourg condition [Ber61]:

$$E_F^c - E_F^v > h\upsilon > E_g,$$

where $E_g$ represents the energy gap of the semiconductor and $E_F^c - E_F^v$ is the energy difference between the quasi-Fermi levels of the conduction band ($E_F^c$) and the valence band ($E_F^v$). This simply states that any photons with energy $h\upsilon$ that fulfill the above condition will be stimulated. Clearly, this system closely resembles the 4-level system underlying the operation of standard lasers based on atomic or molecular transitions. The Bernard–Durrafourg condition is very general and remains valid irrespective of the system's dimensionality, whether for bulk materials and thin films (3D), quantum wells (2D), or nanowires (1D). As shown in Fig. 2-b, gain in a semiconductor can be considered as a 4-level system where stimulated emission occurs near the band edge and pumping occurs from the valence to conduction bands (defining the quasi-Fermi levels), which quickly relax to the band edge (via very fast phonon relaxation) before recombining to emit light.

This can only occur when a strong pulsed laser is used to create many carriers in a very short time, as mentioned previously. In this case, the Fermi level of the semiconductor is not well defined, but we prefer to define quasi-Fermi levels for the conduction and valence bands. These quasi-Fermi levels are equal to the equilibrium Fermi energy when no pulsed light is present or when a continuous-wave laser is used, and is thus not strong enough to deviate from the equilibrium Fermi energy. This is a major difference from standard gas lasers, where the energy levels are intrinsic to the medium used. For semiconductors, one must create this type of 4-level system dynamically by creating these conduction- and valence-band quasi-Fermi levels. Another major difference is that in solids and especially semiconductors, carrier–carrier interaction occurs and may be dominant in some cases, leading to different types of stimulated emission. Slight modifications are necessary depending on the types of carrier involved in the stimulated emission process, whether due to exciton–exciton scattering (so-called P-band) or the formation of an EHP (so-called N-band). As already mentioned, the stimulated emission process in nanostructures is more complex than in standard lasers and has been intensely discussed [Kli07,Kli12]. Nevertheless, for an EHP, the scheme in Fig. 2-b remains valid, except that rather than having the bandgap as the band edge, the bandgap is shifted due to bandgap renormalization for high carrier density above the Mott density. Above this, excitons no longer exist as such, owing to the screening effect of charges, and only a plasma of electrons and holes (N-band) remains [Kli12].

When examining the features of semiconductor nanowire lasers, one must consider the phenomenon of re-absorption, which occurs when photons emitted by the semiconductor material have a high probability to be re-absorbed by the nanostructure. It is particularly strong in semiconductor nanowires for two main reasons. First, the so-called Urbach tail is particularly broad in nanowires [Kli12]. The Urbach tail is present when strong exciton–phonon interaction occurs, and thus 'bends' the bandgap of the material, rendering it less sharp. The other reason is that these structures are good waveguides, confining the light in the structure for longer, therefore increasing the probability of absorption [Li13,Liu13]. Other parameters responsible for the broad Urbach tail, especially in 1D nanostructures, are crystallographic defects and surface states.

Finally, we should mention the important laser feature of coherence. Vanmaekelbergh's group at Utrecht University examined interference effects between light emissions from two ends of a ZnO nanowire up to 20 µm long, with a diameter ranging from 60 to 400 nm. They clearly showed energy spacing between sharp lasing modes scaling with the inverse length of the nanowire [vaV06] (see Table 2), clearly indicating the coherence of the light.

   c. Gain and losses

In the previous section, we derived the intrinsic absorption/gain coefficients for semiconductors, as applied to nanowires. However, this only considered the intrinsic properties of the material and do not account for all the losses. These losses can result from growth conditions giving rise to surface states, but can also be propagation losses of the light within the nanowire, due to tapering or low end-facet reflectivities. In general, the gain for such a system is given by:

$$\Gamma.\gamma = \alpha_{WG} + \alpha_M,$$

where $\Gamma$ is the confinement factor (Table 2), $\alpha_{WG}$ represents the losses due to the propagation of light, and $\alpha_M$ represents the intrinsic losses due to the mirror reflectivities of the nanowire end-facets, as given in Table 2 [Zim10]. The causes of these losses have been addressed above and are a major concern for efficient nanowire laser devices.

To assess the losses due to photonic behaviors in the NW, we must consider the size of the whole nanowire cavity in the Fabry–Pérot regime. Theoretical study and experimental results show that both the length and radius must be greater than 1 μm and 100 nm, respectively, to achieve high $\Gamma$ and reasonable optical feedback. Although different studies of the relevant electromagnetic characteristics of a bare nanowire have demonstrated its ability to behave as a Fabry–Pérot resonator, there are several drawbacks to directly implementing it as monolithic integrated source.

First, the growth mechanism causes the nanowire to either lie on a substrate (horizontal geometry) or stand vertically for surface emission (vertical geometry). The interaction with the substrate in both cases induces severe optical losses, which considerably degrade the Q factor and/or the absolute reflection coefficient at the wire/substrate interface. Furthermore, the assembly of horizontal nanowires on large commercial wafers suggests the need for heavy methods of placement, such as dielectrophoresis, that would not fulfill industrial throughput requirements. For vertical geometry, optical feedback can be reinforced for sufficiently large nanowire diameters with the appearance of high-order whispering gallery–like modes, allowing stimulated emission for a III-V nanopillar on silicon [Chen11]. A proposed alternative design is to insert a thin metallic layer between the substrate and the nanowire, but this substantially complicates fabrication, as wafer bonding and high-aspect-ratio dry etching must be considered [Fri09]. Although this design has promising results, it is unable to take full advantage of the flexibility of bottom-up synthesis.

To reach the laser threshold, the relatively low Q factor (<100) of the NW cavities requires the material gain to compensate for the cavity losses. While this condition is easily realized for II-VI semiconductors and GaN, where a high material gain can be reached ($10^4$ cm$^{-1}$) [Dom96], the large surface-state concentration typically inhibits stimulated emission in III-V nanowires. To date, only two groups have reported stimulated emission in the near-infrared range in a core/shell GaAs/GaAsP bare nanowire [Hua09,Sax13]. Similarly, we must also mention that the inherent longitudinal multimode regime of the Fabry–Pérot resonator, owing to the combination of cavity length and the rather broad linewidth of the wire emission profile at room temperature, is also detrimental to achieving efficient gain in nanowire lasers.

Concerning the lasing properties of these 1D systems, do we observe amplified stimulated emission (ASE) or proper stimulated emission with laser oscillations? The difference between these is that in the first case there is no feedback; ASE occurs simply because many photons are present in the structures, and thus many nearby photogenerated carriers will be stimulated. Proper stimulated emission or lasing emission, meanwhile, requires a feedback mechanism provided by a cavity. In nanowires, a self-formed cavity arises from the end-facets of the nanowire itself, thanks to a strong index mismatch ($n_{NW} \approx 2–3.5$ in semiconductors) with the surrounding environment (usually air). Zimmer *et al.* reported a transition from ASE to lasing in their system [Zim08]. They studied the output intensity as a function of pump power and observed first a linear behavior due to spontaneous emission in ZnO, then a superlinear behavior due to ASE at the edge of the threshold, and finally linear dependence again, indicating that laser oscillation occurs. This 'S' shape in a log–log scale is a clear signature of laser oscillation.

They also found a maximum diameter for laser oscillation to occur, regardless of the nanowire length. This was established in earlier theoretical work [Mas03] analyzing typical laser cavity conditions, showing that, for nanowires, the end-facets of the mirrors are poor reflectors (see Table 1 for cavity properties such as reflectivities and free spectral range-FSR). Indeed, the end-facets of nanowires are determined by their extremities, and the reflection coefficient is simply the nanowire-to-air difference in refraction indices.

To summarize, even though a nanowire laser appears to be an ideal nanosize-laser, in practice, it is not as efficient as desired in current technology. Defects, surface states, multimode lasing, poor end-facet reflectivities, mode overlapping, and mode competition are many aspects that still must be addressed for improved device efficiency. Nanowire lasers are still unable to outdo their thin–film counterparts, where 100–500 kW/cm$^2$ output intensities are typically reported (Table 1) [Wie96,Gad13]. In the next section, we will review how to overcome these losses using advanced photonic designs. Much work remains to be done to unlock the full potential of semiconductor nanowire lasers.

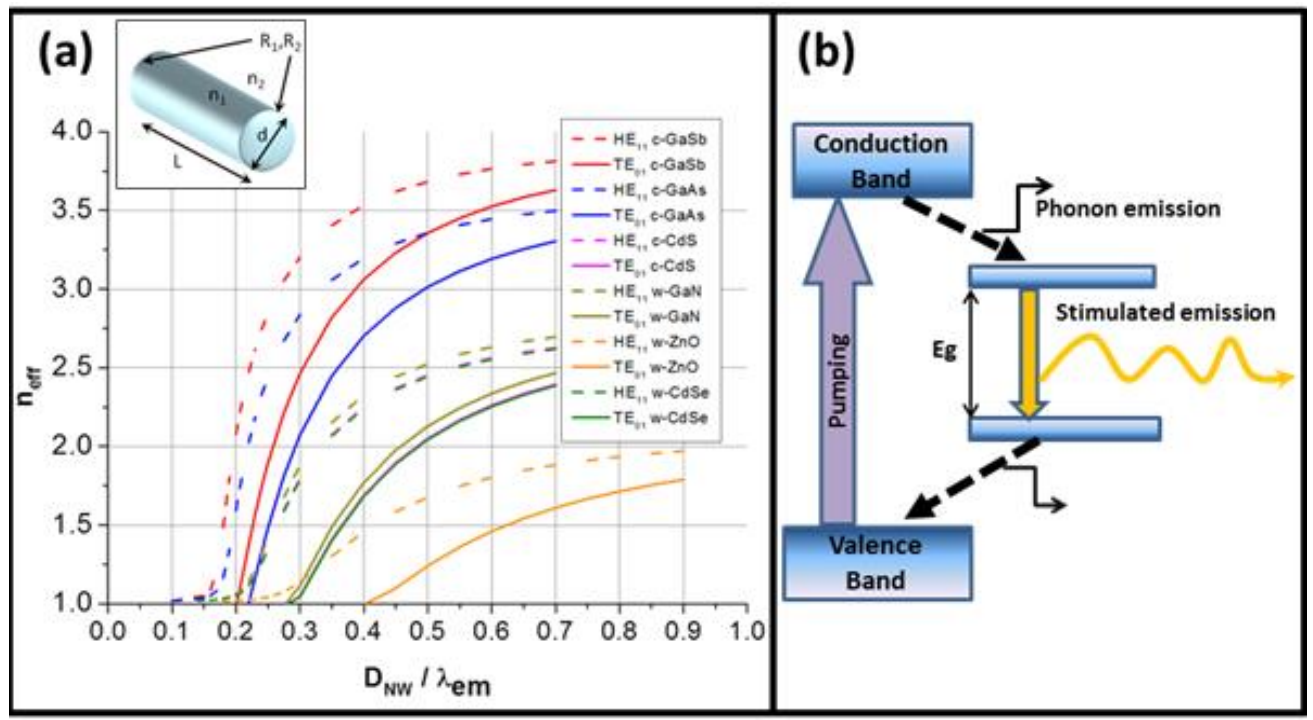

**Figure 2. (a) Effective index of the HE$_{11}$ (dashed) and TE$_{01}$ (solid) propagating modes versus diameter in infinite cylindrical nanowires for different semiconducting materials, with respect to their emission wavelength. The prefix "c" stands for cubic zincblende, and "w" for hexagonal wurtzite crystal phases (Inset: Schematic of a nanowire with relevant parameters: length L, diameter d, internal index $n_1$, medium index $n_2$, and end-facet reflectivities $R_1$ and $R_2$). Index data extracted from Adachi [Ada04] (b) A 4-level system equivalent to a semiconductor.**

**Table 2. Summary of relevant parameters and figures of merit in lasing experiments.**

| Formulae | Meaning |
|---|---|
| $Q = \dfrac{\lambda}{\Delta\lambda} = \dfrac{\nu}{\Delta\nu}$ | Quality factor $Q$ of the cavity |
| $L_c = \dfrac{\lambda_{lnp}}{2}p$ | Cavity resonance condition ($l, n, p$: indices where $p$ is the longitudinal index) |
| $g = \dfrac{1}{2L_c}\ln\dfrac{1}{R^2}$ | Cavity gain ($L_c$: cavity length; $R$: end-facet reflectivity if $R_1 = R_2$) |
| $\nu_F = \dfrac{c}{2nL_c}$ | Free spectral range, spacing of longitudinal modes ($n$: nanowire index) |
| $\Gamma = \dfrac{\int_0^d E^2 dz}{\int_{-\infty}^{\infty} E^2 dz}$ | Confinement factor ($d$: nanowire diameter) |
| $T = T_{max}\dfrac{1}{1 + C\left(\dfrac{\sin(\delta/2)}{\delta/2}\right)^2}$ | Transmission function of a Fabry–Pérot cavity ($\delta$: optical path difference; $C$: contrast) |
| $C = \dfrac{4R}{(1-R)^2}$ | Contrast of a Fabry–Pérot cavity |
| $F = \dfrac{\pi}{2}\sqrt{C} = \dfrac{\sqrt{R}}{1-R}$ | Finesse of a Fabry–Pérot cavity |
| $\tau_c = \dfrac{T_{rt}}{1 - R_1 R_2} \approx \dfrac{2L_C}{\nu_g}\dfrac{1}{1 - R_1 R_2}$ | Photon cavity lifetime ($T_{rt}$: return-trip transmission cavity; $\nu_g$: group velocity) |
| $r_1 r_2 e^{ikL_c} E_0 = E_0$ | Condition of oscillation in a cavity ($r_1, r_2$: complex end-facet reflectivities; $k$: wave-vector; $E_0$: incident EM field) |
| $\lambda_{eff} = \dfrac{\lambda}{n}$ | Effective wavelength of light in a medium with index $n$ |

# 4. Advanced nanowire laser design

As seen above, although semiconductor nanowires have great potential for producing photons, designing an efficient system remains a challenge in practice. As discussed, this is mainly because studies have focused on material issues rather than photonic issues. With better control of growth techniques, it is now possible to seek more advanced laser designs based on existing semiconductor laser devices. In this section, we will review and discuss some avenues of research towards more complex systems that would ultimately be more commercially viable. We begin by reviewing previous work on optoelectrical devices based on NW lasers. We then mention some hybrid systems and heterostructure-based nanolasers before considering photonics and discussing how to achieve single-mode lasing in such structures. Finally, we describe how to engineer photonic structures around a nanowire laser in order to create an efficient and convenient device.

a. Electrical pumping

Duan *et al.* showed that electrically injected carriers in single CdS nanowires could lead to stimulated emission [Dua03]. They achieved the lowest optically pumped single-nanowire lasing threshold of 2 kW/cm$^2$ at low temperature from CdS nanowires grown using laser-assisted catalytic growth. They successfully obtained electrically pumped nanowire lasing around 509 nm by utilizing p-silicon/n-CdS heterojunctions. This was the first optoelectronic device made of nanowire lasers (see Figs. 3-a and b).

Little progress has since been made towards electrically pumped nanowire lasers. Zhang *et al.* showed electrically driven UV lasing behavior from a phosphorus-doped p-ZnO nanonail array/n-Si heterojunction [Zha09b]. Chu *et al.* presented the most striking report of electrical injection in ZnO, by achieving an optically as well as electrically pumped Fabry–Pérot lasing effect from a thin film/nanowire ZnO p-n homojunction (Figs. 3-c and d) [Chu11].

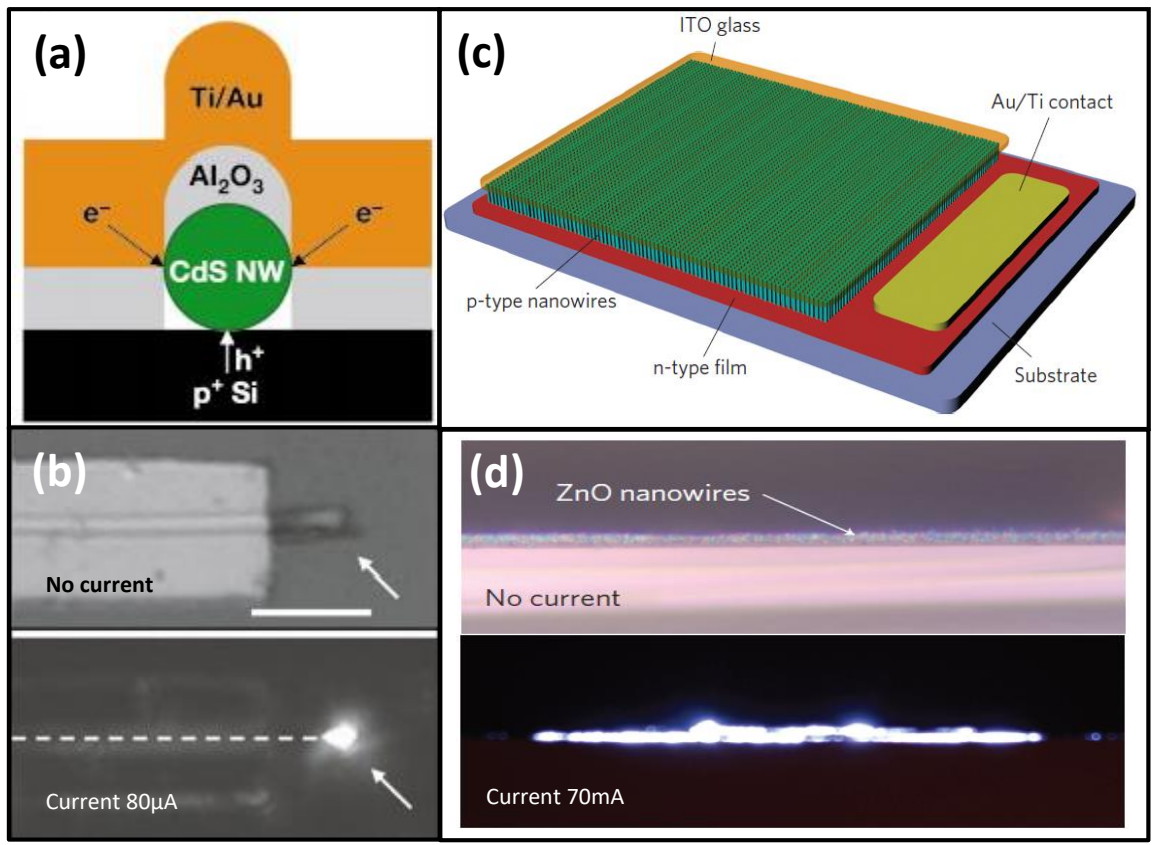

**Figure 3. Electrically driven nanowire lasers made of (left) CdS and (right) ZnO. (a, c) Schematics of the structures [Dua03,Chu11]. (b) SEM images of the device (top) without and (bottom) with current injection (adapted with permission [Dua03]). (d) Side view under optical microscope observation: (top) under illumination and without current injection, (bottom) without illumination and with current injection (adapted with permission [Chu11]).**

For electro-optic devices based on nanowire lasers, Lieber's group also demonstrated electro-optic modulation of CdS and GaN single nanolasers using microfabricated electrodes, which they attributed to an electroabsorption mechanism [Gre05]. This very promising first result has not been seriously followed up on.

Thus far, little work has been done on electrical pumping for lasing applications, as opposed to photovoltaic or photodetection applications. Despite the efforts described above, more remains to be done on electrical injection, especially in III-V materials that have not yet been used as optoelectronic devices for nanolasers, even though they appear the most natural choice considering the success of their quantum well counterparts.

### b- Hybrid systems and heterostructures

When discussing advanced laser design, hybrid systems and heterostructures naturally come to mind. However, like electrical injection, few works have focused on such complex structures. For hybrid systems, one can mix semiconductor nanowires with metallic structures or with dielectric coatings. As mentioned already, we do not intend to describe plasmon-based lasers or spasers, but rather how metal is used to reinforce optical confinement in semiconductor nanowires. Maslov and Ning from the NASA Ames Research Center made the first theoretical study of hybrid systems, where they considered reducing the nanowire diameter while maintaining stimulated emission by coating the nanowire with a metal such as silver [Mas07]. This was reconsidered in an investigation into the modal properties of hybrid plasmonic waveguides [Zhu10]. We must also note that Atwater's group studied the influence of a plasmonic core/shell nanowire resonator on the Purcell factor of a III-V semiconductor nanowire [Hof11]. Bian *et al.* conducted further theoretical studies on nanowire-based hybrid plasmonic structures for low-threshold lasing well below the wavelength scale [Bia13]. Wu *et al.* made one of the first experimental observations of a plasmonic-aided nanolaser and achieved a metal–oxide–semiconductor structure with a bundle of green-emitting InGaN/GaN nanorods strongly coupled to a metal plate through a $SiO_2$ dielectric nanogap layer [Wu11].

Xu *et al.*, meanwhile, presented results of gold substrate-induced single-mode lasing in GaN nanowires [Xu12b]. They showed single-mode lasing and demonstrated that, via the losses due to the gold film, one can discriminate lasing modes such that only one mode will be preferably excited. Similarly, Wu *et al.* developed a hybrid photon–plasmon laser by coupling a CdSe NW with a silver NW. The two nanowires interact over a short length forming a 'X' shape. Careful analysis showed that they coupled photon modes from the CdSe NW to plasmon modes from the Ag NW, thus achieving the first longitudinal coupling between photon and plasmon modes, even allowing for mode selection [Wu13].

Finally, for hybrid systems, there are two reports on nanocables of CdSe and CdS coated with dielectric materials such as $SiO_2$ [Pan05,Ye11]. The CdSe and CdS nanowires were coated with amorphous silica to prevent surface-state recombination and to protect the surface from oxidation and degradation. Moreover, this coating allows better confinement of the electromagnetic mode, as in the case of surface plasmons with metals.

Owing to the difficulty in growing heterostructures from semiconductor nanowires, there are much fewer studies on them than on hybrid systems. In 2008, Qian *et al.* from Lieber's group presented the first nanowire heterostructure exhibiting laser behavior. It consisted of a multi-quantum-well core/shell nanowire [Qia08]. They reported a wide wavelength-tunable laser using a core/shell single nanowire grown on an R-sapphire substrate via metal organic chemical vapor deposition (MOCVD) [Qia08]. Triangular GaN was utilized as the core, and multi-quantum–well (up to 26) InGaN was utilized as the shell. The increase in indium produced a red shift in the laser wavelength. This 'tour de force' is thus far the only one in the literature.

### c- Single-mode selection

Optical signal processing with optical interconnects and photonic integrated circuits requires ultra-compact and low-power integrated lasers providing stable monochromatic light emission. To date, the distributed feedback (DFB) laser remains the preferred choice for most applications, despite its poor capacity for integration with other on-chip optical sources and its relatively large footprint. In terms of emission wavelength range and ease of integration, nanowire lasers can outperform DFB lasers only when laser architectures can provide stable single-mode emission and are compatible with large-scale microfabrication techniques.

Although lasing in nanowires has been experimentally demonstrated as early as in 2001, it was not until 2011 that Xiao *et al.* reported single-mode lasing oscillations within a semiconductor nanowire [Xia11]. By curling and folding a single CdSe nanowire, they observed a reduced number of modes from no loop to a single loop from one end, to two loops from both ends of the nanowire where single mode lasing occurred. Emission occurred at 738 nm with a linewidth of only 0.12 nm and a threshold of 34.4 $\mu J/cm^2$.

There are more ways to obtain a single-mode laser without using an extra external photonic structure. Li *et al.* from Sandia National Labs achieved stable single-mode lasing output from a single GaN nanowire operating far from the lasing threshold [Li12]. A top-down technique exploiting tunable dry etching, combined with anisotropic wet etching for precise control of the wire dimensions, enabled the fabrication of nanowires with a relatively high material gain. They achieved transverse mode selection by utilizing nanowire dimensions that supported a low number of transverse modes within the gain bandwidth. This required significant reduction and precise control of nanowire dimensions, as well as high material gain to compensate for the reduced gain length. Based on a multimode laser theory, they concluded that single-mode lasing arises from strong mode competition and narrow gain bandwidth.

The Vernier effect in a pair of nanowires was used to obtain single-mode lasing. Mode selection was performed and only one mode was stimulated at the end using two GaN nanowires coupled side-by-side [Xu12a]. Yang *et al.* from Berkeley University followed the same principles, except for performing the coupling axially rather than radially [Gao13]. This technique extended the existing concept of a cleaved-coupled cavity (C3) from conventional ridge laser diodes to nanowire lasers. Miniaturization of such devices is a significant technological challenge, and imposes stringent requirements on the air gap dimension between the two Fabry–Pérot cavities and on the crystalline quality of the end-facets of each Fabry–Pérot resonator, in order to achieve reliable axial coupling. In both cases, we actually have two Fabry–Pérot cavities coupled with each other and only one resonant mode in both cavities will be stimulated.

In contrast, Yang *et al.* were the first to couple a ZnO nanowire to a silica microfiber knot cavity [Yan09] with a laser threshold of 0.2 $\mu J/cm^2$. Coupling to the knot cavity allows mode selection, and thus a lower threshold and potentially single-mode lasing. This is also a type of coupled-resonator system, and comparable with the Vernier-effect experiments.

### d- External cavity engineering

This final section focuses on the coupling of nanowires with photonic structures so far. Four photonic structures have been engineered to maximize lasing efficiency: the ring resonator, micro-stadium, Bragg-type structure, and photonic crystals. Figure 4 summarizes the various advanced photonic structures combined with nanowire lasers that have been studied.

Of the different monolithic semiconductor cavities, the ring resonator could initially be seen as incompatible with nanowire geometry. Nevertheless, the formidable mechanical flexibility and strength of semiconductor nanowires combined with advances in micromanipulation allow the development of artificial nanowire "ring" cavities by bending a bare linear nanowire. Pauzauskie *et al.* performed a thorough optical spectroscopy study comparing the spectral properties of a bare GaN nanowire Fabry–Pérot cavity and a GaN nanowire ring resonator [Pau06]. Both structures exhibit lasing oscillations under optical pumping. However, distinct behaviors are clearly observed below and above the threshold. As expected, the nanowire ring resonator exhibits a Q factor close to 1000, which is one order of magnitude higher than for the linear nanowire. Moreover, an emission wavelength shift is observed between the two types of cavities, and the ring resonator favors emissions of longer-wavelength modes via gain discrimination. Following this, Ma *et al.* of the State Key Laboratory, Beijing University, presented similar results, but also showed how to couple it to an artificial ring resonator using another straight CdS nanowire. The Fabry–Pérot fringes were again examined to characterize the system [Ma09].

Once again, Lieber's group was among the first to demonstrate a so-called microstadium nanowire laser, where they coupled a single GaN nanowire within a microresonator in a stadium shape made of $Si_3N_4$ [Par07]. Emission occurred at 372 nm at a relatively high threshold power of 1536 kW/cm$^2$, with a quality factor of 3500. As the refractive index of the $Si_3N_4$ microstadium is less than that of GaN (2.5), light was coupled into the $Si_3N_4$ (with thickness 185 nm), which acted as a microstadium resonator.

To further improve nanowire laser performance, novel nanowire cavities have been proposed, mostly inspired by solid-state optical micro-cavity concepts. Notably, semiconductor microcavities with high Q and small mode volume *V* constitute a promising photonic platform for on-chip low-power nanolasers, or more advanced cavities sustaining strong or weak coupling regimes [Cla10,Hos10]. The design of a monolithic optical microcavity relies on tightly confining light in a small volume using highly reflective photonic mirrors. Recent progress in photonic bandgap materials has aided development of highly reflective monolithic mirrors, such as distributed Bragg mirrors proposed primarily for vertical-cavity surface-emitting lasers (VCSEL). Chen *et al.* from Pittsburgh University introduced an optoelectronic model of a GaN nanolaser with distributed-Bragg-reflector mirrors along a nanowire made of successive AlGaN/GaN layers with lengths 5–10 μm and diameter 120 nm. In particular, they discussed the geometric parameters to achieve single-mode operation conditions [Che06]. However, while this structure is attractive for surface-emitting nanolasers, the incorporation of Bragg mirrors during nanowire growth remains a challenge and imposes stringent constraints on fabrication. Similarly, Zhang *et al.* from Harvard University proposed embedding a horizontal CdS nanowire in a 1D photonic crystal (PhC) cavity formed by PMMA gratings. A Q factor as high as $3\times10^5$ and a mode volume as small as 0.2 $(\lambda/n)^3$ can be achieved in this architecture. Nevertheless, the nanowire diameter must still satisfy the waveguiding conditions discussed previously, which limits the miniaturization of the device [Zha08]. In 2011, Das *et al.* from the University of Michigan proposed an original $SiO_2$ cavity embedding a GaN nanowire as a gain medium and closed by top and bottom $SiO_2/TiO_2$ distributed-Bragg-reflector mirrors. They reported polariton lasing with a record low threshold of 92.5 nJ/cm$^2$ at room temperature, which is at least two orders of magnitude lower than the previously reported polariton laser, and more than three orders of magnitude lower than the conventional photon laser [Das11]. This remarkable result highlights the potential of nanowires as a gain medium for innovative cavity quantum electrodynamics systems.

Barrelet *et al.* made the first report of a photonic device using a single nanowire in 2006, where they coupled a single CdS nanowire to a PhC and a racetrack-type microresonator [Bar06]. PhC cavities are formed by a local perturbation of an otherwise regular pattern of holes (or pillars) in a dielectric structure. Excellent optical resonators have been demonstrated [Son05], featuring both a high Q factor ($> 10^6$) and a very small modal volume ($< (\lambda/n)^3$). PhCs have been considered for providing nanowires with better cavities. By combining the numerous degrees of freedom to design a highly resonant PhC cavity and the flexibility of bandgap engineering during nanowire synthesis, innovative nanolasers can be devised.

Another possibility for combining NWs and PhCs is the direct vertical growth of a nanowire from a specific location on the 2D PhC membrane (Fig. 4). Larrue *et al.* investigated whether this geometry is convenient in terms of optimal light–matter coupling [Lar12]. The Purcell Factor, defining the enhancement of spontaneous emission of a narrow lineshape dipole in a cavity [And99] is a relevant measure; however, it must be remembered that nanowires are not point emitters nor are they narrowband. The reduction of gain at the lasing threshold, or, equivalently, the pump level, is another criterion. This means that a suitable design for the nanowire and PhC cavity can allow a substantial reduction of the lasing threshold of the nanowire and also raise the spontaneous emission factor to ~0.3, which is a typical number for nanolasers [Lar12].

Experimentally, the growth of a nanowire requires a non-conventional alignment (111) of the III-V substrate. Heo *et al.* from the University of Michigan made the first monolithic integration of a single nanowire in a PhC cavity [Heo11]. GaN nanowires were grown on a silicon (111) substrate using plasma-assisted molecular beam epitaxy, and then embedded in a $TiO_2$ PhC slab lying on spin-on glass as a low refractive index material (Fig. 4).

Remarkable progress by IMEC in Belgium in direct selective area growth of III-V compounds on silicon has recently allowed the growth of high-quality InP in narrow $SiO_2$ trenches on silicon (001) substrates by MOCVD [Wan11]. Wang *et al.* analyzed the lasing performances of a micro-laser design similar to that proposed by Larrue *et al.* [Lar12], where a InGaAs/InP pillar was embedded in a silicon PhC cavity [Wan13a,Wan13b]. By overcoming barriers to the epitaxy of high-quality III-V materials on silicon (001) substrates, on-chip lasers formed by integrating an individual nanowire inside a PhC cavity could become a potential building block for low-power integrated sources that have to date been missing from the silicon platform.

A different geometry involves an assembly of horizontal nanowires on top of the PhC layer. This leads to much simpler and stronger light–matter coupling; however, individual nanowires must be placed one-by-one in the cavity [Bir12,Bir13]. This approach was very recently investigated at NTT in Japan. A single InAsP/InP nanowire (emitting around 1.3 µm) was assembled with a slotted PhC cavity on silicon. A narrow photoluminescence peak related to PhC resonance was observed. This approach has many benefits: first, the index matching between silicon and the nanowire is very good, thereby easing the design of a high-Q cavity ($> 10^6$ is expected, nanowire included); second, lasing is potentially very efficient, as the active volume, recipient of the excited carriers, is very small, and could therefore lead to the integration of compact and efficient sources on a silicon platform.

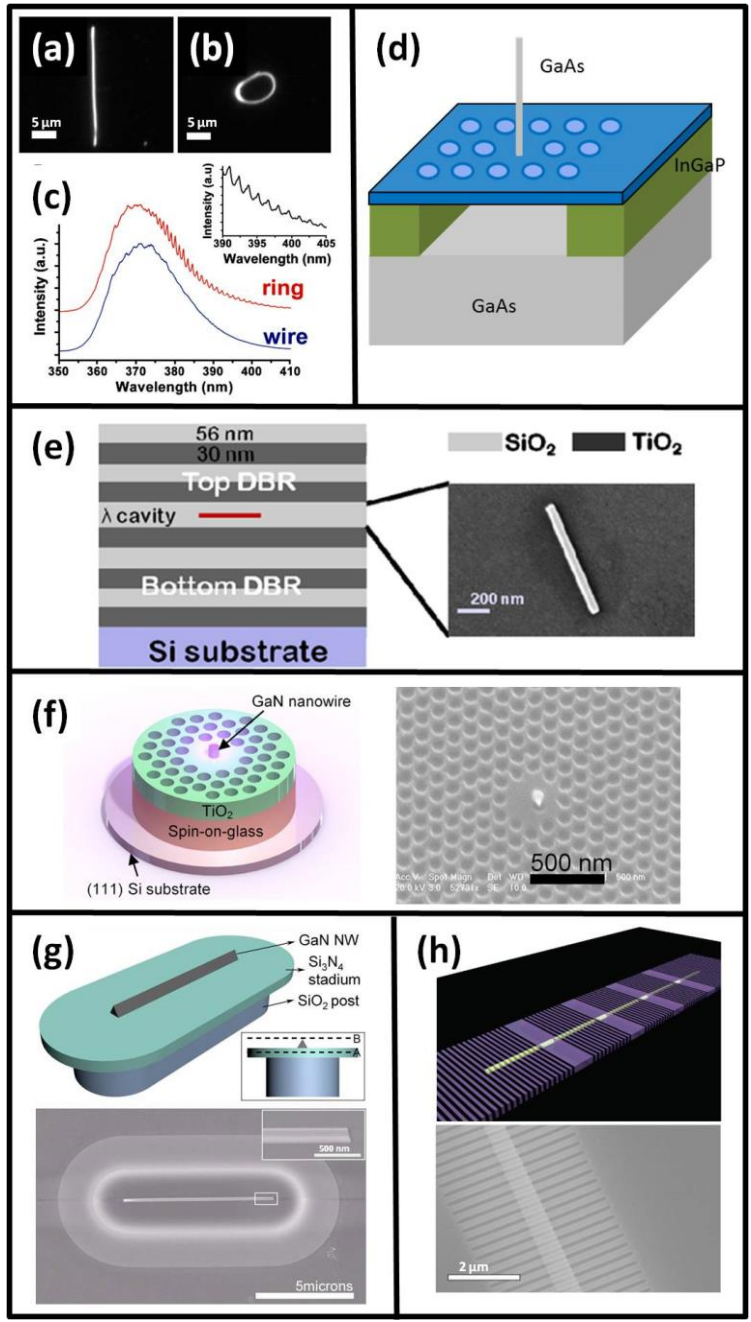

**Figure 4. Various photonic structures.** SEM images of (a) a nanowire and (b) a ring resonator, with (c) their emission spectra (adapted with permission [Pau06]). (d) Vertical nanowire on a PhC cavity (extracted with permission [Lar12]). (e) Left: Schematic of a nanowire in a DBR microcavity; right: SEM image of the nanowire (adapted with permission [Das11]). (f) Schematic (left) and SEM image (right) of a GaN nanowire standing on a PhC (adapted with permission [Heo11]). (g) Schematic (top) and SEM image (bottom) of a nanowire placed on a microstadium structure; inset: focus at the tip of the nanowire (adapted with permission [Par07]). (h) Schematic (top) and SEM image (bottom) of a nanowire lying on a microresonator structure (adapted with permission [Bar06]).

While the incorporation of a single nanowire in a 2D PhC resonator defined by air holes drilled in a semiconductor slab appears very promising, we must mention that vertical nanowire arrays constitute a natural candidate for direct integration of optical components, such as light sources in photonic integrated circuits. Xu *et al.* investigated using 3D FDTD calculations different realistic nanowire array cavities suitable for laser application, and proposed several designs exhibiting mode volumes ranging from ~10 $(\lambda/n)^3$ to 2 $(\lambda/n)^3$ and Q factors as high as $10^4$ [Xu07]. Huffaker's group from UCLA made the first experimental demonstration of a PhC cavity formed by selective area epitaxy of InGaAs/GaAs nanowires [Sco11a]. They then achieved the first bottom-up PhC laser emitting in the near-infrared spectral range. Nanowires formed by axial InGaAs/GaAs heterostructures were first grown on a GaAs (111)B substrate, then removed using a PDMS layer (low index material), thus ensuring an efficient vertical optical confinement. Under optical pumping, fabricated lasers exhibit single-mode emission with a relatively low threshold peak-power density of ~625 $W/cm^2$ [Sco11b]. Figure 4 summarizes what is currently available in the literature regarding complex photonic structures that combine cavities and nanowires for lasing applications.

## 5- Perspectives

Considering the extensive literature and recent innovative experiments involving nanowires, especially the formation of nanowire lasers, it is safe to say that these nanosources of light are very promising as small light sources in future devices for lighting, biomedical, sensing, or energy-harvesting applications.

Nevertheless, more remains to be done for cheap and cost-effective design of such systems. These aspects are improving, but must be better for commercial applications. More must be done on electrical pumping, as only very few groups have tackled the issue of electrically pumped nanowire lasers, but this is essential for commercialization. Finally, packaging is also a major issue when handling such small objects. This has been tackled by several groups, which attempted to couple these nanosources of light with more standard technologies such as optical fibers, but further work is still required. Once these three technological issues are solved, there will be no obstacles against a new generation of nanosources of light being inserted everywhere, e.g. in transparent oxides, in vehicles, and even on clothes and perhaps food for safety and hygiene control.

It is worth mentioning a peculiar feature that may ultimately be very important for practical applications: nanowires can bend easily. The potential of the piezoelectric effect in nanowires, particularly ZnO, is well known [Pan13]. It is worth studying this type of system in order to combine the different features in a single system. Yang *et al.* [Yan13] investigated the bending effect on a CdSe nanowire laser. As the bending radius decreased, they found that the threshold optical pump intensity increased and the differential efficiency decreased. This should be the first of many studies, especially as fundamental optomechanical properties are currently being investigated [Ver12]. Nanowire-based technologies have the potential to combine optical, electrical and mechanical feature all in one go.

## Acknowledgements

We thank A. Gadallah for initial help with the bibliography. C. Couteau would like to acknowledge support from the CNRS PEPS ICQ "InteQ" project and the 'Labex ACTION' program (contract ANR-11-LABX-01-01). A. Larrue thanks the French National Research Agency (ANR) for funding under grant ANR-2011-NANO-028-01. C. Soci acknowledges support from Nanyang Technological University (project reference M4080538 and M4080511) and the Singapore Agency for Science, Technology and Research (A*STAR, SERC Project No.

1223600007). C. Soci and C. Couteau acknowledge support from the Singapore Ministry of Education Academic Research Fund (project reference MOE2011-T3-1-005 and MOE2013-T2-044).